\documentstyle[12pt]{article}
\def\hybrid{\topmargin 0pt      \oddsidemargin 0pt
        \headheight 0pt \headsep 0pt
        \textwidth 17.5cm
        \textheight 25cm
        \voffset=-0.7cm
        \hoffset=-0.4cm
       \hoffset=-1.2cm
        \marginparwidth 0.0in
        \parskip 5pt plus 1pt   \jot = 1.5ex}
\catcode`\@=11
\def\marginnote#1{}

\newcount\hour
\newcount\minute
\newtoks\amorpm
\hour=\time\divide\hour by60
\minute=\time{\multiply\hour by60 \global\advance\minute by-\hour}
\edef\standardtime{{\ifnum\hour<12 \global\amorpm={am}%
        \else\global\amorpm={pm}\advance\hour by-12 \fi
        \ifnum\hour=0 \hour=12 \fi
        \number\hour:\ifnum\minute<10 0\fi\number\minute\the\amorpm}}
\edef\militarytime{\number\hour:\ifnum\minute<10 0\fi\number\minute}

\def\draftlabel#1{{\@bsphack\if@filesw {\let\thepage\relax
   \xdef\@gtempa{\write\@auxout{\string
      \newlabel{#1}{{\@currentlabel}{\thepage}}}}}\@gtempa
   \if@nobreak \ifvmode\nobreak\fi\fi\fi\@esphack}
        \gdef\@eqnlabel{#1}}
\def\@eqnlabel{}
\def\@vacuum{}
\def\draftmarginnote#1{\marginpar{\raggedright\scriptsize\tt#1}}

\def\draft{\oddsidemargin -0.1truein
        \def\@oddfoot{\sl preliminary draft \hfil
        \rm\thepage\hfil\sl\today\quad\militarytime}
        \let\@evenfoot\@oddfoot \overfullrule 3pt
        \let\label=\draftlabel
        \let\marginnote=\draftmarginnote
   \def\@eqnnum{{\rm (\theequation)}\rlap{\kern\marginparsep\tt\@eqnlabel}%
\global\let\@eqnlabel\@vacuum}  }

\newdimen\linethick  \linethick=0.4pt
\newdimen\hboxitspace    \hboxitspace=5pt
\newdimen\vboxitspace    \vboxitspace=5pt

\def\fr#1{%
\beq\new
\vcenter{
\hrule height\linethick
           \hbox{\vrule width\linethick
                 \kern\hboxitspace
                 \vbox{\kern\vboxitspace
                       \hbox{$\begin{array}{c}\displaystyle#1
          \end{array}$}%
                       \kern\vboxitspace}%
                 \kern\hboxitspace
                 \vrule width\linethick}%
           \hrule height\linethick}%
\eeq}

\newdimen\Squaresize \Squaresize=14pt
\newdimen\Thickness \Thickness=0.5pt

\def\Square#1{\hbox{\vrule width \Thickness
   \vbox to \Squaresize{\hrule height \Thickness\vss
      \hbox to \Squaresize{\hss#1\hss}
   \vss\hrule height\Thickness}
\unskip\vrule width \Thickness}
\kern-\Thickness}

\def\Vsquare#1{\vbox{\Square{$#1$}}\kern-\Thickness}

\def\numberbysection{\@addtoreset{equation}{section}
        \def\theequation{\thesection.\arabic{equation}}}
\numberbysection

\renewcommand{\theequation}{\thesection.\arabic{equation}}
\newcommand{\l@qq}[2]{\addvspace{2em}
 \hbox to\textwidth{\hspace{1em}\bf #1 \dotfill #2}}

\newcounter{app}

\def\app{\setcounter{equation}{0}
\def\theequation{\Alph{app}.\arabic{equation}}\par
   \addvspace{4ex}
   \@afterindentfalse
  \secdef\@app\@dapp}
\newcommand\@app{\@startsection {app}{1}{0ex}%
                                   {-3.5ex \@plus -1ex \@minus -.2ex}%
                                   {2.3ex \@plus.2ex}%
                                   {\normalfont\Large\bf}}

\def\@dapp#1{%
{\parindent \z@ \raggedright  \bf #1}\par\nobreak}
\def\l@app#1#2{\ifnum \c@tocdepth >\z@
    \addpenalty\@secpenalty
    \addvspace{1.0em \@plus\p@}%
    \setlength\@tempdima{2.5em}%
    \begingroup
      \parindent \z@ \rightskip \@pnumwidth
      \parfillskip -\@pnumwidth
      \leavevmode \bfseries
      \advance\leftskip\@tempdima
      \hskip -\leftskip
      #1\nobreak\hfil \nobreak\hb@xt@\@pnumwidth{\hss #2}\par
    \endgroup\fi}
\newcounter{sapp}[app]

\def\sapp{\def\theequation{\Alph{app}.\arabic{equation}}\par
   \@afterindentfalse
  \secdef\@sapp\@dsapp}
\newcommand\@sapp{\@startsection{sapp}{2}{\z@}%
                                     {-3.25ex\@plus -1ex \@minus -.2ex}%
                                     {1.5ex \@plus .2ex}%
                                     {\normalfont\large\bfseries}}

\def\@dsapp#1{%
{\parindent \z@ \raggedright  \bf #1}\par\nobreak}
\newcommand{\l@sapp}{\@dottedtocline{2}{1.5em}{3em}}

\def\titlepage{\@restonecolfalse\if@twocolumn\@restonecoltrue\onecolumn
     \else \newpage \fi \thispagestyle{empty}\c@page\z@
        \def\thefootnote{\fnsymbol{footnote}} }

\def\endtitlepage{\if@restonecol\twocolumn \else  \fi
        \def\thefootnote{\arabic{footnote}}
        \setcounter{footnote}{0}}  
\relax

\hybrid

\newtoks\@stequation

\def\subequations{\refstepcounter{equation}%
  \edef\@savedequation{\the\c@equation}%
  \@stequation=\expandafter{\theequation}
  \edef\@savedtheequation{\the\@stequation}
  \edef\oldtheequation{\theequation}%
  \setcounter{equation}{0}%
  \def\theequation{\oldtheequation\alph{equation}}}

\def\endsubequations{%
  \setcounter{equation}{\@savedequation}%
  \@stequation=\expandafter{\@savedtheequation}%
  \edef\theequation{\the\@stequation}%
  \global\@ignoretrue}

\makeatletter
\newdimen\normalarrayskip              
\newdimen\minarrayskip                 
\normalarrayskip\baselineskip
\minarrayskip\jot
\newif\ifold             \oldtrue            \def\new{\oldfalse}
\def\arraymode{\ifold\relax\else\displaystyle\fi} 
\def\eqnumphantom{\phantom{(\theequation)}}     
\def\@arrayskip{\ifold\baselineskip\z@\lineskip\z@
     \else
     \baselineskip\minarrayskip\lineskip1\baselineskip\fi}

\def\@arrayclassz{\ifcase \@lastchclass \@acolampacol \or
\@ampacol \or \or \or \@addamp \or
   \@acolampacol \or \@firstampfalse \@acol \fi
\edef\@preamble{\@preamble
  \ifcase \@chnum
     \hfil$\relax\arraymode\@sharp$\hfil
     \or $\relax\arraymode\@sharp$\hfil
     \or \hfil$\relax\arraymode\@sharp$\fi}}

\def\@array[#1]#2{\setbox\@arstrutbox=\hbox{\vrule
     height\arraystretch \ht\strutbox
     depth\arraystretch \dp\strutbox
     width\z@}\@mkpream{#2}\edef\@preamble{\halign \noexpand\@halignto
\bgroup \tabskip\z@ \@arstrut \@preamble \tabskip\z@ \cr}%
\let\@startpbox\@@startpbox \let\@endpbox\@@endpbox
  \if #1t\vtop \else \if#1b\vbox \else \vcenter \fi\fi
  \bgroup \let\par\relax
  \let\@sharp##\let\protect\relax
  \@arrayskip\@preamble}
\def\eqnarray{\stepcounter{equation}%
              \let\@currentlabel=\theequation
              \global\@eqnswtrue
              \global\@eqcnt\z@
              \tabskip\@centering                      
              \let\\=\@eqncr
              $$%
            \halign to \displaywidth  \bgroup
             \eqnumphantom \@eqnsel
      \hskip\@centering                               
    $\displaystyle  \tabskip\z@ {##}$%
    &\global\@eqcnt\@ne \hskip 2\arraycolsep
         $ \displaystyle  \arraymode{##}$\hfil
    &\global\@eqcnt\tw@ \hskip 2\arraycolsep
         $\displaystyle\tabskip\z@{##}$\hfil
         \tabskip\@centering
    &{##}\tabskip\z@\cr}
\makeatother

\def\bea{\begin{eqnarray}}
\def\eea{\end{eqnarray}}

\def\beq{\begin{equation}}
\def\eeq{\end{equation}}
\def\be{\beq\new\begin{array}{c}}
\def\ee{\end{array}\eeq}
\def\bse{\begin{subequations}}                
\def\ese{\end{subequations}}                 %

\begin{document}
\vspace{0.2cm}
\begin{center}
{\LARGE \bf The Stringy Representation of } \\
\vspace{0.5cm}{\LARGE \bf the $D\geq{3}$ Yang-Mills Theory. } \\
\vspace{0.7cm} {\large\bf Andrey Yu. Dubin}\\
\vspace{0.5cm}{\bf ITEP, B.Cheremushkinskaya 25, Moscow 117259, Russia}\\
\vspace{0.5cm}{{\it e-mail: dubin@heron.itep.ru}}\\
{tel.:~7 095 129 9674~~~~  ;  ~~~~fax:~7 095 883 9601}
 \end{center}
\vspace{0.3cm}

\begin{abstract}

I put forward the stringy representation of the $1/N$ strong coupling ($SC$)
expansion for the regularized Wilson's loop-averages in the {\it continuous}
$D\geq{3}$ Yang-Mills theory ($YM_{D}$) with a sufficiently large {\it bare}
coupling constant $\lambda>\lambda_{cr}$ and a fixed ultraviolet cut off
$\Lambda$. The proposed representation is proved to provide with the
{\it confining} solution of the Dyson-Schwinger chain of the judiciously
{\it regularized} $U(N)$ Loop equations. Building on
the results obtained, we suggest the stringy pattern of the low-energy
theory associated to the $D=4$ $U(\infty)\cong{SU(\infty)}$ gauge theory in
the standard $\lambda\rightarrow{0}$ phase with the asymptotic freedom in the
$UV$ domain. A nontrivial test, to clarify whether the $AdS/CFT$
correspondence conjecture may be indeed applicable to the large $N$ pure
$YM_{4}$ theory in the $\lambda\rightarrow{\infty}$ limit, is also discussed.

\end{abstract}

\begin{center}
\vspace{0.5cm}{Keywords: Yang-Mills, Loop equation, Duality, String,
Strong-coupling expansion} \\
\vspace{0.2cm}{PACS codes 11.15.Pg; 11.15.Me; 12.38.Aw; 12.38.Lg}
\end{center}

\newpage

\section{Introduction.}

There is some evidence \cite{Wils,'t Hooft1} that the $D\geq{3}$ pure
Yang-Mills systems ($YM_{D}$), including the conventional one with the action
\be
S=\frac{1}{4g^{2}}\int d^{D}x~tr\left( F_{\mu\nu}(x)F_{\mu\nu}(x)\right)~,
\label{1.1}
\ee
might be reformulated as a sort of string theory. Recently, synthesizing
the nonabelian duality transformation \cite{Dub2} with the Gross-Taylor
stringy reformulation \cite{Gr&Tayl} of the $D=2$ $YM_{2}$ theory,
a concrete proposal \cite{Dub3} has been made for the stringy representation
of the $1/N$ strong coupling ($SC$) expansion in a generic continuous
$D\geq{3}$ $YM_{D}$ system with a fixed ultraviolet ($UV$) cut off $\Lambda$.
The construction is built on the key-observation revealing the
exact $WC/SC$ {\it correspondence} which takes place (prior to the $UV$
regularization) between the two alternative $1/N$ series.
Certain conglomerates of the Feynman diagrams, comprising the $1/N$
weak-coupling ($WC$) series in a given continuous gauge system, via a formal
resummation can be traded for the judiciously associated
variety of the appropriately weighted (piecewise) smooth worldsheets of the
color-electric flux, facilitating the $1/N$ $SC$ expansion in the same
$YM_{D}$ system.

Given the proper regularization (that introduces a {\it nonzero} width of
the $YM$ vortex), a particular variant of the latter $SC$
expansion is supposed to faithfully represent, at least for sufficiently
large $N$, the regularized $U(N)$ gauge theory (\ref{1.1}) provided that 
the dimensionless and $N$-{\it independent} {\it bare} coupling constant 
\be
\lambda=(g^{2}N)\Lambda^{D-4}>\lambda_{cr}(D)
\label{1.1bx}
\ee
is {\it larger} than certain critical value $\lambda_{cr}(D)$ presumably
associated to the large $N$ phase transition. For the final justification
of the asserted $YM_{D}/String$ duality, in the present paper I follow the
somewhat complementary root. Given the $U(N)$ theory (\ref{1.1}), our aim is
to solve the Dyson-Schwinger chain of the loop equations \cite{LE/MM,LE/P}
which, in the $N\rightarrow{\infty}$ limit, reduces to the single Loop
equation
\be
\hat{\mathcal{L}}_{\nu}({\bf x}(s))<W_{C}>_{\infty}=\tilde{g}^{2}
\oint\limits_{C} dy_{\nu}(s')~\delta_{D}({\bf y}(s')-{\bf x}(s))
<W_{C_{xy}}>_{\infty}<W_{C_{yx}}>_{\infty},
\label{1.11b}
\ee
where $\tilde{g}^{2}=g^{2}N=\lambda\Lambda^{4-D}$,
$\hat{\mathcal{L}}_{\nu}({\bf x})$ is the Loop operator specified
by eq. (\ref{1.6bf}) below, while $<W_{C}>_{\infty}$ denotes the
$N\rightarrow{\infty}$ limit of the correlator of the Wilson
loop\footnote{Owing to the constraint imposed by the $D$-dimensional
$\delta_{D}(..)$-function, the r.h. side of eq. (\ref{1.11b}) vanishes unless
$C$ has a selfintersection at a point ${\bf x}(s)={\bf y}(s')$ so that
$C\equiv{C_{xx}}=C_{xy}C_{yx}$ is decomposable into the two subloops $C_{xy}$
and $C_{yx}$.}
\be
W_{C}=\frac{1}{N}Tr\Bigg[~{\mathcal{P}}~exp\left(i\oint_{C} dx_{\mu}A_{\mu}(x)
\right)\Bigg]~~~~~,~~~~~A_{\mu}(x)\equiv{A^{a}_{\mu}(x)T^{a}_{ij}}~.
\label{1.2}
\ee
It can be then demonstrated, see \cite{Dub4}, that the derived $D\geq{3}$
{\it confining} solution (of the loop equations) matches, to all orders of
the $1/N$ expansion, the appropriately regularized implementation of the
Gauge String proposed in \cite{Dub3}.

The subtlety is that, in the $SC$ phase (\ref{1.1bx}), the transverse profile
of the {\it microscopic} flux-tubes does depend on a particular choice of the prescription to
implement the $UV$ regularization. Therefore, to attack the loop equations
in the economic way, it is vital to reduce the regularization-dependence of
the $YM_{D}$ loop-average $<W_{C}>$ as much as possible.
Upon a reflection, there are {\it two} limiting regimes (formalized by eqs.
(\ref{0.1eea}) and (\ref{0.1eed}) below) where the dependence of $<W_{C}>$ on
the choice of the regularization indeed can be reduced to the dependence of a
few relevant coupling constants, entering the corresponding formulation
of the Gauge String representation, on the bare coupling (\ref{1.1bx}).
In both cases, we
are dealing with the dominance of the {\it infrared} phenomena: the contours
$C$, being constrained to possess the radius of curvature ${\mathcal{R}}(s)>>
\Lambda^{-1}$ (for $\forall{s}$), should be associated to
sufficiently large values of the minimal area $A_{min}(C)$ of the
saddle-point worldsheet $\tilde{M}_{min}(C)$ spanned by $C$. Given the latter
conditions, the required reduction is shown to be maintained when the
characteristic amplitude $\sqrt{<{\bf h^{2}}>}$ of the worldsheet's
fluctuations is {\it either} much larger
\be
\frac{<{\bf h^{2}}>}{<{\bf r^{2}}>}\sim
\frac{D-2}{\lambda}\cdot ln[A_{min}(C)\Lambda^{2}]~\longrightarrow{~\infty}~~~~,
~~~~{\mathcal{R}}(s)\Lambda~\longrightarrow{~\infty}~,
\label{0.1eea}
\ee
{\it or} much smaller
\be
\frac{<{\bf h^{2}}>}{<{\bf r^{2}}>}\sim\frac{ln[A_{min}(C)\Lambda^{2}]}{\lambda}~\longrightarrow{~0}~~~,
~~~[A_{min}(C)\Lambda^{2}]~\longrightarrow{~\infty}~~~,~~~
{\mathcal{R}}(s)\Lambda~\longrightarrow{~\infty}~,
\label{0.1eed}
\ee
than the flux-tube's width $\sqrt{<{\bf r^{2}}>}\sim{\Lambda^{-1}}$.

In both of the above regimes, the corresponding solution of the
Dyson-Schwinger chain allows to reconstruct the associated regularization
of Gauge String representation \cite{Dub3} (of the Wilson's loop-averages).
The prescription is that the 'bare' worldsheet's weight (assigned to the
infinitely thin flux-tubes) of \cite{Dub3} is substituted by its smeared
counterpart, visualized through the {\it fat} $YM$ vortices, so that the
nonuniversality of the smearing is essentially unobservable under the
considered conditions (\ref{0.1eea}) or (\ref{0.1eed}). The discussed solution
is expected to describe {\it stable} stringy excitations only in the $SC$
phase (\ref{1.1bx}). At least for sufficiently large $N$, this is
predetermined by the fact that, within the $1/N$ $SC$ expansion, the physical
string tension $\sigma_{ph}$  mandatory exhibits the $\Lambda^{2}$-scaling
(\ref{0.1eec}). Nevertheless, combining the present results with the outcome
of \cite{Dub3}, we suggest the stringy pattern of the low-energy theory
associated to the $D=4$ $U(\infty)\cong{SU(\infty)}$ gauge theory
(\ref{1.1}) in the conventional $WC$ phase $\lambda\rightarrow{0}$.

\section{The regime (\ref{0.1eea}) of the large fluctuations.}

The complexity of eq. (\ref{1.11b}) is foreshadowed by its {\it nonlinearity}
that takes place when the loop $C$ nontrivially selfintersects.
Complementary, one may expect that, in eq. (\ref{1.11b}), {\it not} any
'natural' (from the $\lambda\rightarrow{0}$ phase viewpoint) regularization
of the $\delta_{D}({\bf x}-{\bf y})$-function can be translated into a
{\it tractable} regularization of the presumable stringy solution of the Loop
equation in the $SC$ phase. On the contrary, in the sector $\Upsilon_{0}$ (of
the full loop space $\Upsilon$) comprised of the contours $C$ {\it without}
nontrivial selfintersections, eq. (\ref{1.11b}) considerably simplifies so
that a simple gauge-invariant regularization results in the
transparent pattern of the regularized stringy solution.
To take advantage of the latter simplification, the idea is to handle
the Loop equation (\ref{1.11b}) in the two steps. First, one is
to find a subclass of the regularized solutions of the considered
$\Upsilon_{0}$-reduction of the Loop equation. Then, we find out
under what circumstances thus obtained solutions correctly reproduce the
loop-averages for nontrivially {\it selfintersecting} contours.

\subsection{The solutions of the Loop Equation on $\Upsilon_{0}$.}

For any nonintersecting loop $C\in{\Upsilon_{0}}$, the
r.h. side of eq. (\ref{1.11b}) receives a nonzero contribution only from the
{\it trivial} selfintersection point ${\bf x}(s)={\bf y}(s')$ with ${s'=s}$
so that one can put $<W_{C_{xy}}>=<W_{C_{xx}}>$ while $<W_{C_{yx}}>=1$.
Actually, the same simplification takes place for the finite $N$ extension
of eq. (\ref{1.11b}) so that the latter finite $N$ equation is reduced on
$\Upsilon_{0}$ to the {\it linear} one
\be                                                      
\hat{\mathcal{L}}_{\nu}({\bf x}(s))<W_{C}>=\tilde{g}^{2}<W_{C}>
\oint\limits_{C} dy_{\nu}(s')~\delta_{D}({\bf y}(s')-{\bf x}(s))~~~~,~~~~
C\in{\Upsilon_{0}}~,
\label{0.9za}
\ee
where $\tilde{g}^{2}=g^{2}N\sim{N^{0}}$, and one can implement (consistently
with the manifest gauge invariance) the smearing prescription
\be
\delta_{D}({\bf x}-{\bf y})~\longrightarrow{~
{\Lambda}^{D}{\mathcal{G}}({\Lambda}^{2}({\bf x}-{\bf y})^{2})}
~~~~~;~~~~~\int d^{D}z~{\mathcal{G}}({\bf z}^{2})=1~,
\label{0.1bb}
\ee
where ${\mathcal{G}}({\bf z}^{2})$ is a sufficiently smooth function (so that
all its moments are well-defined) which satisifies the natural
normalization-condition.

Next, a priori, one can search for the solution of the reduced eq.
(\ref{0.9za}) in the form of the regularized sum
\be
<W_{C}>={\sum_{\tilde{M}}}^{(r)}~w_{2}[\tilde{M}(C)]
\label{0.1bbj}
\ee
over the worldsheets $\tilde{M}(C)$ weighted by a factor
$w_{2}[\tilde{M}(C)]$. Akin to the conventional Nambu-Goto
theory, the surfaces $\tilde{M}(C)$ are supposed to result from the smooth
immersions into the Euclidean base-space ${\bf R^{D}}$, and
the sum's superscript $(r)$ recalls about
the $UV$ cut off $\Lambda$ for the transverse fluctuations of the string.
Then, plunging the stringy Ansatz (\ref{0.1bbj})
into eq. (\ref{0.9za}), the latter equation can be transformed into the one
(to be regularized according to eq. (\ref{0.1bb}))
\be
\hat{\mathcal{L}}_{\nu}({\bf x}(s))~
ln\left(w_{2}[\tilde{M}(C)]\right)=\tilde{g}^{2}
\oint\limits_{C} dy_{\nu}(s')~\delta_{D}({\bf y}(s')-{\bf x}(s))
\label{1.11bfb}
\ee
which operates directly with the worldsheet's weight $w_{2}[\tilde{M}(C)]$.
In going over from eq. (\ref{0.9za}) to eq. (\ref{1.11bfb}), we have taken
into account that the Loop operator\footnote{
In eq. (\ref{1.11bfb}), $\delta/\delta\sigma_{\mu\nu}({\bf x}(s))$ and
$\partial_{\mu}^{{\bf x}(s)}$ denote respectively the Mandelstam
area-derivative and the path-derivative \cite{LE/MM}.}
\be
\hat{\mathcal{L}}_{\nu}({\bf x}(s))=
\partial_{\mu}^{{\bf x}(s)}~
\frac{\delta}{\delta \sigma_{\mu\nu}({\bf x}(s))}
\label{1.6bf}
\ee
complies with the {\it Leibnitz} rule. Due to the first-order nature of
$\hat{\mathcal{L}}_{\nu}$, the general solution $w_{2}[\tilde{M}(C)]$ of eq.
(\ref{1.11bfb}) assumes the form
\be
w_{2}[\tilde{M}(C)]=\tilde{w}_{2}[\tilde{M}(C)]~w^{(0)}_{2}[\tilde{M}(C)]
~~~~~~~;~~~~~~~
\hat{\mathcal{L}}_{\nu}({\bf x}(s))~ln(w^{(0)}_{2}[\tilde{M}(C)])=0,
\label{1.9add}
\ee
where $\tilde{w}_{2}[..]$ is any particular solution of eq. (\ref{1.11bfb}),
while $w^{(0)}_{2}[\tilde{M}(C)]$ is formally allowed to be an arbitrary
$N$-{\it independent zero mode} (of the Loop operator) fulfilling some
natural cluster decomposition requirement.

As it is demonstrated in \cite{Dub4} (where the general solution of the
zero-mode equation (\ref{1.9add}) is found), the zero-mode factor does not
alter the picture arising when one puts $w^{(0)}_{2}[\tilde{M}(C)]=1$
retaining only the appropriate particular solution $\tilde{w}_{2}[..]$.
To derive the latter, it is helpful to utilize the abelian Stokes theorem and
rewrite the r.h. side of eq. (\ref{1.11bfb}) as the surface integral over
$\tilde{M}(C)$. Then, after some simple
manipulations, one can reduce (see \cite{Dub4} for the details) eq.
(\ref{1.11bfb}) to the transparent equation
\be
\frac{2~\delta^{2}~ln\left(\tilde{w}_{2}[\tilde{M}(C)]\right)}
{\delta\sigma_{\mu\nu}({\bf x}(s))~\delta\sigma_{\rho\chi}({\bf y}(s'))}=
-(\delta_{\mu\rho}\delta_{\nu\chi}-
\delta_{\mu\chi}\delta_{\nu\rho})~\lambda{\Lambda}^{4}~
{\mathcal{G}}({\Lambda}^{2}({\bf x}-{\bf y})^{2})~,
\label{1.11bb}
\ee
Finally, the structure of eq. (\ref{1.11bb}) is suggestive that the Mandelstam
area-derivatives might be traded, $\delta/\delta \sigma_{\mu\nu}({\bf x}(s))
\rightarrow{\delta_{f}/\delta p_{\mu\nu}({\bf x}(\gamma))
|_{{\bf x}(\gamma)={\bf x}(s)\in{C}}}$, for the ordinary
functional area-derivatives (preliminary restricted to the boundary $C$)
with respect to the standard infinitesimal area-element
$d\sigma_{\mu\nu}({\bf x}(\gamma))=p_{\mu\nu}(\gamma)~d^{2}\gamma,~
p_{\mu\nu}(\gamma)=\varepsilon^{ab}~\partial_{a} x_{\mu}(\gamma)
\partial_{b} x_{\nu}(\gamma)$, 
where the coordinates $x_{\mu}(\gamma)\equiv{x_{\mu}(\gamma_{1},\gamma_{2})}$
define the position of a given worldsheet $\tilde{M}$ in the base-space
${\bf R^{D}}$. The considered substitution is supported by the pattern of the
resulting solution\footnote{In fact, the pattern (\ref{0.1}) is reminiscent
of (but {\it not} equivalent to) the {\it ad hoc} smearing \cite{LE/MM} of
the $m_{0}=0$ option of the Nambu-Goto weight (\ref{2.5bb}). Complementary,
the weight (\ref{0.1}) can be viewed as a specific implementation of the
general confining string Ansatz \cite{PolyakCS} which, in turn, is routed in
the abelian Kalb-Ramond pattern \cite{Kalb&Ramond}.}
\be
\tilde{w}_{2}[\tilde{M}_{\chi}]=N^{\chi}~
exp\left(-\frac{\lambda\Lambda^{2}}{4}
\int\limits_{\tilde{M}_{\chi}}\int\limits_{\tilde{M}_{\chi}}
d\sigma_{\mu\nu}({\bf x})d\sigma_{\mu\nu}({\bf y})~
{\Lambda}^{2}{\mathcal{G}}({\Lambda}^{2}({\bf x}-{\bf y})^{2})
\right)~,
\label{0.1}          
\ee
where, advancing ahead, we have included the 't Hooft topological factor
(with $\chi$ being the total Euler character of $\tilde{M}_{\chi}$) which
justification requires to deal with the full chain of the
loop equations.

Actually, the quasi-local pattern (\ref{0.1}) can be linked,
despite its somewhat bizzare appearence, to the realm of the conventional
stringy models. For this purpose, we employ the following {\it infrared
universality} taking place in the regime
(\ref{0.1eea}) as far as macroscopic contours without zig-zag backtrackings\footnote{
The prescription, to implement the mandatory backtracking invariance of the
$YM_{D}$ loop-averages $<W_{C}>$, is given in \cite{Dub4}.} are concerned.
The key-observation \cite{Dub4} is that, in this case, {\it the above
solution of the Loop equation is supposed to be in the one infrared
universality class with the unconventional (owing to eq. (\ref{0.1eec})
below) implementation of the Nambu-Goto string} which, in
turn, is supposed to be reformulated in the spirit of the
'low-energy' noncritical Polyakov's theory. Thus associated
Nambu-Goto theory, presumed to possess the same {\it UV} cut off $\Lambda$,
is endowed with the weight
\be                            
w_{1}[\tilde{M}_{\chi}]=N^{\chi}~
exp\left({-\frac{\bar{\lambda}(\lambda)\cdot\Lambda^{2}}{2}
A[\tilde{M}_{\chi}]}-m_{0}(\lambda)L[\tilde{M}_{\chi}]\right)~,
\label{2.5bb}
\ee
where $L[\tilde{M}]$ is the length of the boundary $\partial{\tilde{M}}$ of
$\tilde{M}$, while $\bar{\lambda}(\lambda)$ and $m_{0}(\lambda)$ are
certain functions of $\lambda$ which depend on the choice of {\it both} the
flux-tube's transverse profile {\it and} the prescription for
the regularization of the string fluctuations. (More refined analysis
\cite{Dub4} demonstrates that the simple $L[\tilde{M}_{\chi}]$-dependence of
the subleading boundary-contribution is valid for $<W_{C}>$ only provided that
all the loop's {\it self}intersections, if present, are {\it point-like} from
the low-energy viewpoint.) Let us also remark that the validity of the
considered infrared equivalence implies that the contour $C$ is
{\it macroscopic} in the concrete sense of the following twofold constraint.
Firstly, except for the $1/\Lambda$-vicinity of the nontrivial (point-like)
selfintersections of $C$, the distance
$|{\bf x}(s)-{\bf x}(s')|$ is much larger than the flux-tube's width
$\sqrt{<{\bf r^{2}}>}\sim{\Lambda^{-1}}$,
provided that the length of the corresponding segment of the
boundary $C=\partial \tilde{M}$ is much larger than $\Lambda^{-1}$:
\be
\int_{s'}^{s} dt~\sqrt{\dot{x}^{2}_{\mu}(t)}~>>\Lambda^{-1}~~~~~
\Longrightarrow~~~~~{|{\bf x}(s)-{\bf x}(s')|>>\Lambda^{-1}}~.
\label{0.1eel}
\ee
Secondly, we always presume that, once $C_{xx}=C_{xy}C_{yx}$ (where
${\bf x}(s)={\bf y}(s'),~s\neq{s'}$) satisfies the above
condition, then both $C_{xy}$ and $C_{yx}$ comply with this condition as well.
It is also noteworthy that the pattern (\ref{2.5bb}) allows to make a direct
contact \cite{Dub4} with the proposal of \cite{Dub3}.

\subsection{The consistency with the full Dyson-Schwinger chain.}

Let us now turn to the restrictions which are necessary to impose, for the
(multiloop generalization of the) solution (\ref{0.1bbj})/(\ref{1.9add}) to be
consistent with large $N$ Loop equation (\ref{1.11b}) (and, more generally,
with the full chain of the $U(N)$ loop equations). We refer the reder to
\cite{Dub4} for the details and here simply present the outcome.

To begin with, one observes that in the $N=1$ case the reduced eq.
(\ref{0.9za}) becomes the {\it exact} $U(1)$ Loop equation valid
for an {\it arbitrary} selfintersecting contour $C$.
Furthermore, one can easily verify that the $N=1$ Ansatz
(\ref{0.1bbj})/(\ref{1.9add}) is the exact solution of the entire abelian
Dyson-Schwinger chain. The subtle point is that the considered stringy system
describes, for $g^{2}>g^{2}_{cr}$, the electrodynamics enriched with the
monopoles (similarly to the abelian systems discussed in \cite{PolyakCS}) in
the $SC$ phase where the latter are supposed to condense. Another nontrivial
point is that the choice of a particular zero-mode solution
$w^{(0)}_{2}[\tilde{M}(C)]$
is tantamount to the particular choice of the monopole's action.

Next, consider the case of sufficiently large $N$ when the Loop equation
(\ref{1.11b}) is the adequate leading approximation. Then, there are three
major constraints to ensure that there is such a judicious regularization of
the full eq. (\ref{1.11b}) (and of the entire chain) which makes the latter
equation consistent with the $\Upsilon_{0}$-solution
(\ref{0.1bbj})/(\ref{1.9add}). First of all, one has to impose the conditions
(\ref{0.1eea}) which maintain that the characteristic amplitude of the
string's fluctuations (estimated by the same token as in \cite{Luscher}) is
much larger than the vortex width $\sqrt{<{\bf r^{2}}>}\sim{\Lambda^{-1}}$. In
this way, we render unobservable the quasi-contact interactions (additional to
the flux-tube's selfenergy) between the elementary $YM$ vortices. In turn, the
necessity of this suppression is foreshadowed by the abelian nature of the
latter interactions as it is clear from the above discussion of the $N=1$
case. Secondly, in the regime (\ref{0.1eea}), the consistency with
the Dyson-Schwinger chain unambiguously fixes the $N$-dependence of the
$U(N)$ solution in the form of the 't Hooft factor
$N^{2-2h-b}$ (remaining undetermined from eq. (\ref{0.9za}) alone).
Actually, for the number of the boundary loops $b\geq{2}$
and large $N\neq{\infty}$, one is to impose the stricter condition to 
suppress as $e^{-\beta N}$ (with some $\beta>0$) all kinds of the
quasi-contact interactions. Both the characteristic size (understood in the
sense of eq. (\ref{0.1eel})) of each loop, presumed to be devoid of $1d$
selfintersections, and the minimal distance, between any two different
contours, should be of order of $N^{\alpha}$ with some $\alpha>0$.

The third condition implies that, despite its familiar appearence, the system
(\ref{0.1bbj})/(\ref{2.5bb}) is {\it not} entirely conventional at least for
sufficiently large $N\geq{2}$. Within the $1/N$ $SC$ expansion, for the
consistent regularization of the Loop equation (\ref{1.11b}) to exist, the
confining solution should mandatory result in {\it the physical string tension
$\sigma_{ph}$ which is of order of (or, when $\lambda\rightarrow{\infty}$,
much larger than) the {\it UV} cut off $\Lambda$ squared},
\be
\Lambda^{2}=O(\sigma_{ph})~~~~~~~~;~~~~~~~~~
\sigma^{(sc)}_{ph}=\left(\frac{\bar{\lambda}(\lambda)}{2}-\zeta_{D}
\right)\Lambda^{2}~.
\label{0.1eec}
\ee
Similarly to \cite{Luscher}, in the $N\rightarrow{\infty}$ semiclassical
approximation $\sigma^{(sc)}_{ph}$ for $\sigma_{ph}$, the $D\geq{3}$ entropy
contribution $\delta \sigma_{ent}=-\zeta_{D}\Lambda^{2}$ (due to
the regularized transverse string fluctuations) is represented by
the $\lambda$-{\it independent} constant $\xi_{D}$ computed in \cite{Dub4}. 
In particular, eq. (\ref{0.1eec}) implies that, motivating the constraint
(\ref{1.1bx}), the coupling $\bar{\lambda}(\lambda)$ in eq. (\ref{2.5bb}) is
constrained to be sufficiently large.

At first glance, the $\Lambda^{2}$-scaling\footnote{It is this scaling that
hinders the direct application of the considered construction to the
$\lambda\rightarrow{0}$ regime.} (\ref{0.1eec}) looks like
a rather unnatural condition. On the second thought, one could expect this
constraint beforehand: the strongly coupled $D=4$ $YM_{4}$ system (\ref{1.1})
can be reinterpreted as a local {\it prototype} \cite{Dub3} of the effective
low-energy theory for the gauge system (\ref{1.1}) in the
$\lambda\rightarrow{0}$ regime with the
asymptotic freedom in the $UV$ domain. In this  perspective, in eq.
(\ref{0.1eec}) the cut off $\Lambda$ is to be identified with the
confinement-scale which (in the $D=4$ system (\ref{1.1}) with
$\lambda\rightarrow{0}$ and the new $UV$ cut off $\bar{\Lambda}>>\Lambda$) is
supposed to be of order of the lowest glueball mass. From the results
obtained, the considered confinement-scale can be complementary viewed as the
scale where the logarithmic renormgroup flow (of the running coupling
constant) freezes.

Furthermore, according to the arguments of \cite{Dub4}, the
$\Lambda^{2}$-scaling (\ref{0.1eec}) is sufficient for the system
(\ref{0.1bbj})/(\ref{2.5bb}) to suppress the outgrowth of the microscopic
{\it baby-universes} so that the Gauge String avoids the branched polymer
phase. This matches the observation that the full-fledged
quantum analysis of the stringy system (\ref{0.1bbj})/(\ref{2.5bb})
requires to reformulate the latter as a somewhat unconventional string
theory as it suggested by the concept of the Pauli-Villars regularization.
Possessing the $UV$ cut off $\tilde{\Lambda}>>
\sqrt{\sigma_{ph}}\sim{\Lambda}$, this theory should not exhibit propagating
degrees of freedom at the 'short distance' scales $<<{1/\sqrt{\sigma_{ph}}}$,
while approaching the proposed implementation (\ref{0.1eec}) of the Nambu-Goto
pattern (\ref{0.1bbj})/(\ref{2.5bb}) at the scales larger than
$1/\sqrt{\sigma_{ph}}$.

\section{The regime (\ref{0.1eed}) of the quasi-static $YM$ vortex.}

The land-mark of the so far considered regime (\ref{0.1eea}) is that the
quasi-contact interactions between the elementary $YM$ vortices,
being discarded by the sheer Nambu-Goto pattern (\ref{2.5bb}),
are unobservable. In fact, it is tantamount \cite{Dub3,Dub4} to the
unobservability of the choice of the particular action and nonabelian group
defining a given dual gauge theory. The latter choice can be traced back most
transparently in the extreme $SC$ limit (\ref{0.1eed}) where the
$YM_{D}\rightarrow{YM_{2}}$ dimensional reduction (see eq. (\ref{0.5bxx})
below), anticipated in \cite{Dub3}, takes place. To reveal this
phenomenon in the simplest setting, one is to take
advantage of the fact that the $N=1$ option of the Ansatz
(\ref{0.1bbj})/(\ref{0.1}) correctly reproduces the $SC$ expansion in the
$D\geq{2}$ $U(1)$ gauge theory (\ref{1.1}) modified by the presence of the
monopoles in $D\geq{3}$. In particular, the latter $N=1$ Ansatz is applicable
not only to the regime (\ref{0.1eea}) but also to the extreme $SC$ limit
(\ref{0.1eed}). As a result, the abelian analysis provides with the following
general motivations presumably valid for $\forall{N}\geq{1}$.

We proceed by noticing that, in the limit (\ref{0.1eed}), the leading
asymptotics of the average $<W_{C}>$ is supposed to be given by the
contribution of the saddle-point $YM$ vortex corresponding
to the minumal area worldsheet(s) $\tilde{M}_{min}(C)$ with the
characteristic radius of curvature
\be
{\mathcal{R}}(\gamma)\Lambda~\longrightarrow{~\infty}~,
\label{0.1eeb}
\ee
i.e. being (at any point $\gamma\equiv{(\gamma_{1},\gamma_{2})}$ of the
surface) much larger than the flux-tube's width $\sim{\Lambda^{-1}}$.
Let the saddle-point (minimal area) worldsheet(s) $\tilde{M}_{min}(C)$ possess
in ${\bf R^{D}}$ the support $T_{min}=T_{min}(C)$ which, for simplicity, is
presumed to be
{\it unique} for any given $C$ in question. Also, to avoid certain technical
complications, we will presume that $T_{min}(C)$ can be embedded into a
$2d$ manifold ${\bf B^{2}}$ diffeomorphic to ${\bf R^{2}}$.
Then, the $SC$ asymptotics (\ref{0.1eed}) of the $N=1$ Ansatz
(\ref{0.1bbj})/(\ref{0.1}) can be written as the $N=1$ option of the general
pattern
\be
<W_{C}>\Bigg|_{YM_{2}(T_{min})}=\sum_{\{R_{q}\}}
F(\{R_{q}\})~exp\left(-\frac{\xi\lambda\Lambda^{2}}{2}
\sum_{q}~C_{2}(R_{q})~\bar{A}_{q}\right)
\label{2.5bxh}
\ee
where the parameter $\xi$ is defined by eq. (\ref{0.1ew}) below,
$C_{2}(R)$ is the eigenvalue of the second $U(N)$ Casimir operator
associated to the representations $R$ (so
that $C_{2}(n)=n^{2}$ in the $U(1)$ case), while $\bar{A}_{q}$ is the area
$A[T_{q}]$ of the $q$th window $T_{q}$ of the support ${T_{min}(C)}$. As for
the origin of $T_{q}$, the support $T_{min}(C)$ is devided into the union
$\{T_{q};~\sum_{q} \bar{A}_{q}=A[T_{min}(C)]\}$ of the disjoint windows
$T_{q}$ after cutting along the loop $C$. Remark
that each domain $T_{q}(\{C_{k}\})$ is assigned with some representation
$R_{q}$ (entering $C_{2}(R_{q})$) of the Lie group in
question, and the sum in eq. (\ref{2.5bxh}) runs over all admissible
$\{R_{q}\}$-assignements satisfying the (nonabelian) {\it fusion-rule}
algebra. The latter is imposed by the $\{\bar{A}_{q}\}$-independent function
$F(\{R_{q}\})$.

Upon a reflection, eq. (\ref{2.5bxh}) yields the average
$<W_{C}>|_{YM_{2}(T_{min})}$ evaluated in the continuous $D=2$
$YM_{2}$ theory (\ref{1.1}) conventionally defined on
$T_{min}(C)$ as on the base-space. Altogether, it motivates the following
statement (substantiated in \cite{Dub3,Dub4} in the alternative way). Keeping
the conditions (\ref{0.1eed})
fulfilled, let both the coupling constant $\lambda$ and
$({\mathcal{R}}(s)\Lambda)$ be much larger than $N^{2}$. Given the above
conditions and summing\footnote{When both $\lambda$ and
$({\mathcal{R}}(s)\Lambda)$ are large but $\leq{N^{2}}$, in
the r.h. side of eq. (\ref{0.5bxx}) one is to retain only the leading $O(N)$
order of the $1/N$ expansion of the $YM_{2}(T_{min})$ average.} up the leading
(with respect to the $1/\lambda$- and
$1/(<{\mathcal{R}}(\gamma)>\Lambda)-$expansions) subseries in $1/N^{2}$,
the pattern of the $D\geq{3}$ $YM_{D}$ averages $<W_{C}>|_{YM_{D}}$ is
supposed to exhibit the reduction
\be
<W_{C}>|_{YM_{D}}~\longrightarrow{~<W_{C}>|_{YM_{2}(T_{min})}}~~~~~~if
~~~~~~\lambda,~({\mathcal{R}}(s)\Lambda)~~>>N^{2}~,
\label{0.5bxx}
\ee
where the $YM_{2}$ coupling constant $g_{YM_{2}}$ is related to the original
$D\geq{3}$ $YM_{D}$ constant $g_{YM_{D}}$ via the rescaling 
\be
g^{2}_{YM_{2}}=\xi~g^{2}_{YM_{D}}~\Lambda^{D-2}=
\xi~\frac{\lambda}{N}~\Lambda^{2}
\label{0.4bc}
\ee
that can be interpreted as the $YM_{D}\rightarrow{YM_{2}}$ {\it dimensional
reduction} implemented modulo the auxiliary $\xi$-factor. In turn,
the parameter\footnote{Furthermore, one can argue that (at least when
$\lambda\sim{1}$) the factor $\xi$, reflecting the $D\geq{3}$ regularization
ambiguity, is {\it of order of unity} once the solution (\ref{0.1}) describes
stable, rather than metastable, stringy excitations.} $\xi$, reflecting the
ambiguity of the $UV$ regularization,
is related to the smearing function (\ref{0.1bb}),
\be
\xi=\int d^{2}z~{\mathcal{G}}({\bf z}^{2})\sim{1}~~~~~~~~;~~~~~~~~
\sigma_{0}=\frac{\xi\lambda\Lambda^{2}}{2}~.
\label{0.1ew}
\ee
that defines the corresponding bare string tension $\sigma_{0}$. Below, we
will prove the large $N$ variant of the $SC$ asymptotics (\ref{0.5bxx})
directly from the Loop equation (\ref{1.11b}), while now it is appropriate to
discuss some implications of eq. (\ref{0.5bxx}).

To begin with, the $YM_{2}$-average in the r.h. side of eq. (\ref{0.5bxx})
can be reformulated in the manifestly stringy terms employing the
construction of \cite{Gr&Tayl,Dub3}. To draw a particularly transparent
consequence of
the $SC$ asymptotics (\ref{0.5bxx}), consider the average $<W^{R}_{C'}>$ of
the Wilson loop in any {\it (anti)chiral} representation $R\in{Y^{(N)}_{n}}$
of the $U(N)$ group. Then, eq. (\ref{0.5bxx}) implies that the leading
asymptotics of the physical string tension (associated to $<W^{R}_{C'}>$) is
proportional,
\be
\sigma_{ph}(R)~\longrightarrow{~\frac{C_{2}(R)}{N}~\sigma_{0}}~~~~~~~;~~~~~~~
R\in{Y^{(N)}_{n}}~,
\label{0.1eww}
\ee
to the eigenvalue $C_{2}(R)$ of the second $U(N)$ Casimir operator.
Then, to compare this pattern with the classification of
the (dual) abelian superconductors, we observe that $C_{2}(R)\geq{|n(R)|}$,
where $n(R)=\sum_{i=1}^{N} n_{i}$ measures the total amount of the elementary
$U(N)$ fluxes. (Conventionally, $N$ nondecreasing integers $n_{i}$
parametrize a generic $U(N)$ representation $R$ so that
$C_{2}(R)\sim{N}$ once $|n(R)|\sim{N^{0}}$.) As a result, the averaged force
between the collinear elementary $YM$ vortices is {\it repulsion}. The Gauge
String, associated to the gauge theory (\ref{1.1}), corresponds therefore to
the type-II superconductor.

To make contact with the $N\geq{2}$ Ansatz (\ref{0.1bbj})/(\ref{0.1}), remark
first that one obtains $\tilde{\sigma}_{ph}(R)\rightarrow
{|n(R)|\sigma_{0}}$, when $\lambda\rightarrow{\infty}$ in the regime
(\ref{0.1eea}). Complementary, the $SC$ asymptotics (\ref{0.1eed}) of the
$N\geq{2}$ system (\ref{0.1bbj})/(\ref{0.1}) complies with eq. (\ref{0.5bxx})
once the saddle-point flux-tube is essentially nonselfoverlapping. In turn,
it ensures that the quasi-contact interactions (between the elementary $YM$
vortices) are unobservable in this case which, in the leading order, leaves
the sheer $m_{0}=0,~\chi=1$ Nambu-Goto pattern (\ref{2.5bb}) applied to the
minimal area surface $\tilde{M}_{min}(C)$.

\subsection{Sketching the derivation of the $SC$ asymptotics (\ref{0.5bxx}).}

For simplicity, we restrict our discussion to the
$N\rightarrow{\infty}$ case imposing, as previously, that the support
$T_{min}(C)$ of the 'minimal area' flux-tube (spanned by $C$) can be
embedded into a curved two-dimensional space ${\bf B^{2}}$ diffeomorphic
to the $2d$ plane ${\bf R^{2}}$. In this case, the generally covariant
extension of the $2d$ $YM_{2}$ theory (\ref{1.1}) can be introduced on
${\bf B^{2}}$ in the standard way so that the pattern of the r.h. side of eq.
(\ref{0.5bxx}) can be borrowed from the ${\bf B^{2}}={\bf R^{2}}$ computations
\cite{Kaz&Kost}. Let a judicious smearing $<W_{C}>|_{YM_{2}(T_{min})}
\longrightarrow{<W_{C}>|^{(reg)}_{YM_{2}(T_{min})}}$,
\be
<W_{C}>|^{(reg)}_{YM_{2}(T_{min})}\Bigg/<W_{C}>|_{YM_{2}(T_{min})}~~=~~
1+O\left((<{\mathcal{R}}(s)>\Lambda)^{-\alpha}\right)~~~~,~~~~\alpha>0~,
\label{0.3bbj}
\ee
is unobservable in the regime (\ref{0.1eed}). The statement is that thus 
regularized r.h. side of eq. (\ref{0.5bxx}) satisfies the (regularized)
$D$-dimensional
Loop equation (\ref{1.11b}). The key-point of the proof is that, when the
considered pattern is plunged (in the regime (\ref{0.1eed})) into eq.
(\ref{1.11b}), the latter equation is reduced to the {\it two-}dimensional
Loop equation conventionally associated to the $D=2$ $YM_{2}$ theory
(\ref{1.1})/(\ref{0.4bc}) on ${\bf B^{2}}$. In turn, the resulting $D=2$ Loop
equation is fulfilled by the very construction (\ref{0.3bbj}) of the smearing
for the $SC$ asymptotics (\ref{0.5bxx}).

To more explicit, in eq. (\ref{2.5bxh}), the suitable smearing prescription
is to substitute each particular $2A_{q}$ by the $\tilde{M}\rightarrow{T_{q}}$
option of the bilocal integral (\ref{0.1}). Being understood in the
stronger sense of eq. (\ref{0.5bxx}) (complemented by the constraint
$\bar{A}_{q}\Lambda^{2}>>N^{4}$, for $\forall{q}$, together with the
two more requirements formulated below), the condition
(\ref{0.1eeb}) justifies the formal substitution
\be
{\Lambda}^{2}{\mathcal{G}}({\Lambda}^{2}({\bf x}-{\bf y})^{2})
\longrightarrow{\xi~\delta^{w}_{2}({\bf x}(\gamma)-{\bf y}(\gamma'))}~,
\label{0.3b}          
\ee
where $\delta^{w}_{2}({\bf x}-{\bf y})$ is the 2-dimensional delta-function
on $T_{min}(C)$, and $\xi$ is defined in eq. (\ref{0.1ew}). One
is to require also that (at any point of $T_{min}(C)$) the line in the normal
direction either does not have the second intersection with $T_{min}(C)$ or
the second intersection takes place at a distance $>>\Lambda^{-1}$.
Additionally, for the validity of eq. (\ref{0.3bbj}), we have to ensure the
stricter suppression of the next-to-leading boundary contribution,
proportional to the effective perimeter-mass $m_{0}(\lambda)$ (which
dependence on the smearing function (\ref{0.1bb}) is evaluated in
\cite{Dub4}), that introduces an extra constraint on
${\mathcal{G}}({\bf z}^{2})$. Altogether, the imposed conditions allow to
substantiate eq. (\ref{0.3bbj}).

Next, to justify the asserted dimensional reduction of the Loop equation
(\ref{1.11b}), one is to synthesize the following two observations.
Firstly, the above smearing prescription together with the condition
(\ref{0.1eeb}) ensure that, acting on the smeared pattern (\ref{2.5bxh}), the
$D$-dimensional Loop operator $\hat{\mathcal{L}}^{\nu}({\bf x}(s))$  reduces
to the {\it two}-dimensional one
\be
\hat{\mathcal{L}}^{\alpha}({\bf x}(s))=\frac{1}{\sqrt{\hat{g}({\bf x})}}~
\partial_{\beta}^{{\bf x}(s)}~\sqrt{\hat{g}({\bf x})}~
\hat{g}^{\beta\gamma}({\bf x})~
\hat{g}^{\alpha\lambda}({\bf x})~
\frac{\delta}{\delta \sigma^{\gamma\lambda}({\bf x}(s))}
\label{1.6bff}
\ee
where $\hat{g}_{\alpha\beta}({\bf x})$ is the $2d$ metric on ${\bf B^{2}}$,
and $\hat{g}({\bf x})\equiv{det_{\{\alpha\beta\}}
[\hat{g}_{\alpha\beta}({\bf x})]}$.
The asserted reduction\footnote{Similar reduction, but in the context of the
$\lambda\rightarrow{0}$ $YM_{D}$ theory (\ref{0.1}) considered {\it prior} to
the $UV$ regularization, is discussed in \cite{Ol&Pet}.} takes place due to
the evident property of the 2-tensor ${\mathcal{K}}_{\mu\nu}({\bf x}(s))$,
resulting after the action of the '$D$-dimensional' Mandelstam derivative
$\delta/\delta \sigma^{\mu\nu}({\bf x}(s))$ onto the smeared
exponent of eq. (\ref{2.5bxh}). The constraint (\ref{0.1eeb}) implies that 
the tensor ${\mathcal{K}}_{\mu\nu}({\bf x}(s))$ has nonvanishing components
only in the subspace spanned by the two Zweibein $D$-vectors
${\bf e}^{\beta}({\bf x})\in{\bf t_{x}B^{2}}$ beloning to the tangent space
${\bf t_{x}B^{2}}$ of ${\bf B^{2}}$ at a given point ${\bf x}={\bf x}(s)$
so that $({\bf e}^{\alpha}({\bf x})|{\bf e}^{\beta}({\bf x}))\equiv{
\sum_{\mu}e_{\mu}^{\alpha}({\bf x})e_{\mu}^{\beta}({\bf x})}=
\hat{g}^{\alpha\beta}({\bf x})$.

The second observation is that, being devided by $\Lambda^{D-2}$,
the regularization (\ref{0.1bb}) of the $D$-dimensional $\delta_{D}$-function
(in the r.h. side of the properly regularized eq. (\ref{1.11b}))
simultaneously can be reinterpreted as the smearing (\ref{0.3b}) of the
$2$-dimensional $\xi\delta_{2}(..)$-function on ${\bf B^{2}}$, where $\xi$ is
defined by eq. (\ref{0.1ew}). (Complementary, the value of
$\xi\tilde{g}^{2}\Lambda^{D-2}$ is equal to the coupling constant
(\ref{0.4bc})
in the $YM_{2}$ theory entering the asymptotics (\ref{0.5bxx}).)
In turn, the regularized $\delta_{2}(..)$-function refers to the r.h. side
of the Loop equation conventionally associated to the generally
covariant extension of the $2d$ $YM_{2}$ theory (\ref{1.1})/(\ref{0.4bc})
defined on the $2d$ manifold ${\bf B^{2}}$ into which the support
$T_{min}(C)$ is embedded. As for the economic regularization of the r.h. side
of eq. (\ref{1.11b}), the following simple prescription is sufficient in the
regime (\ref{0.1eed}) provided the relevant contours are presumed to be 
{\it macroscopic} (according to the definition given prior to eq.
(\ref{0.1eel})). In this case, one can introduce the regularization
{\it separately} for the $1/\Lambda$-vicinities of the nontrivial (point-like)
selfintersections of $C$ (i.e. when $s'\neq{s}$) and for the remaining 
trivial selfintersections (i.e. when $s'={s}$).
Namely, consider for simplicity a macroscopic contour $C$ with a single
nontrivial selfintersection at ${\bf x}(s_{1})={\bf y}(s_{2})$. For
${\bf x}(s)={\bf x}(s_{1})$, the r.h. side of eq. (\ref{1.11b}) is given by
the weighted sum $a_{1}<W_{C_{xx}}>_{\infty}+
a_{2}<W_{C_{xy}}>_{\infty}<W_{C_{yx}}>_{\infty}$ where the
weights $a_{1},a_{2}$ can be represented by the judiciously selected portions
\cite{Kaz&Kost} of the contour integral along $C$. Then, the regularization
prescription is to perform the substitution (\ref{0.1bb}) separately within
$a_{1}$ and $a_{2}$.

Summarizing, after the proper smearing (\ref{0.3bbj}), the Ansatz
(\ref{0.5bxx}) indeed reduces thus regularized $D$-dimensional eq.
(\ref{1.11b}) to the regularized two-dimensional Loop equation defined on
${\bf B^{2}}$. To finally put all the pieces together, one can show that, for
the macroscopic loops in the considered regime (\ref{0.1eed}), the
regularization of the latter $2d$ equation is {\it unobservable} that happily
matches the complementary unobservability (\ref{0.3bbj}).
This matching implies that thus regularized
$2d$ Loop equation merges\footnote{The only subtlety is due to certain extra
'anomaly term' arising when the original variant of area-derivative (with the 
area-variation $|\delta\sigma_{\mu\nu}({\bf x})|<<\Lambda^{-1}$) is to be
transformed into the one with
$|\delta\sigma_{\mu\nu}({\bf x})|>>\tilde{\Lambda}^{-1}$. In our computation, this
perimeter-dependent anomaly can be independently recovered comparing
the action of the area-derivative on the bilocal pattern (\ref{0.1}) in
both  of the relevant limiting cases.} with the equation \cite{Kaz&Kost}
derived for the $YM_{2}$ theory (\ref{1.1})/(\ref{0.4bc})  
with the {\it infinite} (in the units of $(g^{2}_{YM_{2}}N)^{1/2}$)
$UV$ cut off $\tilde{\Lambda}\rightarrow{\infty}$ which,
in turn, justifies the properly smeared asymptotics (\ref{0.5bxx}) as the
solution of eq. (\ref{1.11b}).

\section{Conclusions.}

Hopefully, the present analysis can be included, as a building block, to a
larger scheme which would make accessible the solution of the $D=4$ Loop
equation (\ref{1.11b}) in the most interesting $\lambda\rightarrow{0}$ phase
of the $YM_{4}$ theory (\ref{1.1}) (with the asymptotic freedom in the $UV$
domain). The promising sign is that the derived $YM_{D}\rightarrow{YM_{2}}$
dimensional reduction (\ref{0.5bxx}) is reminiscent of the one in the model
\cite{AmbOles} which is rather popular for the approximation of the lattice
results presumably associated to the gauge theory (\ref{1.1}). Complementary,
consider
the low-energy dynamics of the $U(\infty)$ or $SU(\infty)$ $YM_{4}$ theory
(\ref{1.1}) (with $\lambda\rightarrow{0}$ and the new $UV$ cut off
$\bar{\Lambda}>>\Lambda$) at the confinement-scale where the logarithmic
renormgroup flow, valid in the $UV$ domain, is presumably freezed.
Then, the latter infrared dynamics is supposed to be correctly described by
the $N\rightarrow{\infty}$ limit of the proposed implementation
(\ref{0.1eec}) of the Nambu-Goto system (\ref{0.1bbj})/(\ref{2.5bb}),
provided we are dealing with such correlators where the quasi-contact
interactions of the fat $YM$ vortices are unobservable. This statement is
supported by the observation \cite{Dub4} that the large $N$ limit of the
considered Nambu-Goto sum is expected to be {\it common} for the infrared
description of {\it any} $D\geq{3}$ $U(\infty)$ or $SU(\infty)$ pure gauge
system with an arbitrary polynomial (in terms of $F_{\mu\nu}$) lagrangian
\cite{Dub3} providing with the $N^{2}$-scaling of the free energy.

Finally, it is noteworthy that the regime (\ref{0.1eed}) is
reminiscent of the one which independently appeared within the alternative
stringy description \cite{Witt2} (see also \cite{Gr&Ooguri}) of the pure
gauge theories which is based on the conjecture \cite{Maldac} about the
so-called {\it AdS/CFT correspondence}. Given any macroscopic
nonselfintersecting contour, in the extreme $D=4$ {\it SC} regime
$\lambda>>1$, the purported $\lambda\Lambda^{2}$-scaling \cite{Witt2} of the
physical string tension is consistent with the prediction \cite{Dub3} of our
formalism. For an arbitrarily selfintersecting loop, the predictions
of \cite{Witt2} are still missing. The pattern of the $SC$ asymptotics
(\ref{0.5bxx}), being actually valid for a generic $YM_{D}$ system with
the lagrangian polynomial in term of $F_{\mu\nu}$, imposes the concrete test
(to be compared with the arguments of \cite{Gr&Ooguri})
on the applicability of \cite{Witt2}.

\begin{center}
{\bf Acknowledgements.}
\end{center}

The author is grateful to Yu.Makeenko for comments at the final stage of
the work. This project is partially supported by CRDF grant RP1-2108.

\enddocument